\documentclass[prb,groupedaddress,nofootinbib,noshowpacs,twocolumn,floatfix,preprintnumbers,draft,citeautoscript]{revtex4}

\usepackage[T1]{fontenc}
\usepackage{amsmath}

\def\dotp#1#2{
    #1 \cdot #2}

\newcommand{\ygg}{\ensuremath{\mathrm{U(1)}_Y}}

\newcommand{\bphi}{\mbox{\boldmath $\phi$}}

\newcommand{\half}{\frac{1}{2}}

\newcommand{\gev}{\hbox{ GeV}}

\newcommand{\tev}{\hbox{ TeV}}

\newcommand{\cm}{\hbox{ cm}}

\newcommand{\lum}{\hbox{ cm}^{-2}\hbox{ s}^{-1}}
\newcommand{\eqn}[1]{(\ref{#1})}
\newcommand{\Eqn}[1]{Eq.~(\ref{#1})}
\newcommand{\abs}[1]{\left| #1\right|}

\newcommand{\ewgg}{\ensuremath{\mathrm{SU(2)_L}\otimes \mathrm{U(1)}_Y}}

\newcommand{\wigg}{\ensuremath{\mathrm{SU(2)_L}}}
\newcommand{\isogg}{\ensuremath{\mathrm{SU(2)}}}
\newcommand{\emgg}{\ensuremath{\mathrm{U(1)}_{\mathrm{em}}}}

\newcommand{\D}{\ensuremath{\mathcal{D}}}
\newcommand{\btau}{\mbox{\boldmath $\tau$}}

\newcommand{\balpha}{\mbox{\boldmath $\alpha$}}

\def\tr#1{\mathrm{tr}\!\left(#1\right)}

\newcommand{\gf}{\ensuremath{G_{\mathrm{F}}}}
\newcommand{\lag}{\ensuremath{\mathcal{L}}}

\def\vev#1{\left\langle #1\right\rangle_0}

\def\bentarrow{\:\raisebox{1.3ex}{\rlap{$\vert$}}\!\rightarrow}

\def\dknuc#1#2#3{
	\begin{equation}
	\begin{array}{r c l}
	#1 & \rightarrow & #2 \\
	 & & \phantom{^{152}\;}\bentarrow #3
	\end{array}
	\end{equation}
		}

\begin{document}

\preprint{FERMILAB--PUB--15/058--T}
\title{Electroweak Symmetry Breaking in Historical Perspective}

\author{Chris Quigg}
\email[E-mail address: ]{quigg@fnal.gov}
\affiliation{Theoretical Physics Department\\
Fermi National Accelerator Laboratory\\
P.O. Box 500, Batavia, Illinois 60510 USA}

\begin{abstract}
The discovery of  the  Higgs boson is a major milestone in our progress toward understanding  the natural world. A particular aim of this article is to show how diverse ideas came together in the conception of electroweak symmetry breaking that led up to the discovery. I will also survey what we know that we did not know before, what properties of the Higgs boson remain to be established, and what new questions we may now hope to address. 
\end{abstract}

\maketitle


\section{INTRODUCTION \label{sec:Intro}}
A lively  continuing conversation between experiment and theory has brought us to a  radically simple conception of the material world. Fundamental particles called quarks and leptons are the stuff of direct experience, and two new laws of nature govern their interactions.  Pursuing clues from experiment, theorists have constructed the electroweak theory~\cite{Glashow:1961tr,Weinberg:1967tq,Salam} and quantum chromodynamics~\cite{Fritzsch:1973pi,Gross:1973id,Politzer:1973fx,Kronfeld:2010bx}, refined them within the framework of local gauge symmetries, and elaborated their consequences. In the electroweak theory, electromagnetism and the weak interactions---so different in  range and apparent strength---are ascribed to a common gauge symmetry. We say that the electroweak gauge symmetry is broken, by dynamics or circumstances, to the  gauge symmetry of electromagnetism. 

The electroweak theory and  quantum chromodynamics (QCD) join to form the \textit{standard model} of particle physics. Augmented to incorporate neutrino masses and lepton mixing, the standard model describes a vast array of experimental information. The gauge theories of the strong, weak, and electromagnetic interactions have been validated by experiment to an extraordinary degree as relativistic quantum field theories. Recent textbook treatments of QCD and the electroweak theory may be found, for example in Refs.~ \onlinecite{Langacker:2010,GT2,MattS,DynSM}.

Until recently, the triumph of this new picture has been incomplete, notably because we had not identified the agent that differentiates electromagnetism from the weak interaction. The 2012 discovery of the Higgs boson by the ATLAS~\cite{Aad:2012tfa} and CMS~\cite{Chatrchyan:2012ufa} Collaborations working at CERN's Large Hadron Collider capped a four-decades-long quest for that agent. [Further details of the discoveries are reported in Refs. ~\onlinecite{DellaNegra21122012,CMSCollaboration21122012,ATLASCollaboration21122012,doi:10.1146/annurev-nucl-102313-025603}.] The observations indicate that the electroweak symmetry is \textit{spontaneously broken,} or \textit{hidden}: the vacuum state does not exhibit the full symmetry on which the theory is founded.  Crucial insights into spontaneously broken gauge theories were developed a half-century ago by Englert \& Brout~\cite{Englert:1964et}, Higgs~\cite{Higgs:1964ia,Higgs:1964pj}, and Guralnik, Hagen, \& Kibble~\cite{Guralnik:1964eu}. All the experimental information we have~\cite{CMSHresults,ATLASHresults,Agashe:2014kda,HiggsPDG14} tells us that the unstable 125-GeV particle discovered in the LHC experiments behaves like an elementary scalar consistent with the properties anticipated for the standard-model Higgs boson.

The first goal of this article is to sketch how a broad range of concepts, drawn mainly from weak-interaction phenomenology, gauge field theories, and condensed-matter physics, came together in the electroweak theory. The presentation complements the construction of the electroweak theory given in my pre-discovery article, ``Unanswered Questions in the Electroweak Theory''~\cite{Quigg:2009vq}.  Presentations similar in spirit may be found in Refs. ~\onlinecite{Weinberg:2008zzb,WilczekMIT}. Next, I will briefly summarize what we now know about the Higgs boson, what the discovery has taught us, and why the discovery is important to our conception of nature. Finally, I will address what remains to find out about the 125-GeV Higgs boson and what new questions are opened by its existence. For example, 
we need  to discover what accounts for the masses of the electron and the other leptons and quarks, without which there would be no atoms, no chemistry, no liquids or solids---no stable structures. In the standard electroweak theory, both tasks are the work of the Higgs boson. Moreover, we have reason to believe that the electroweak theory is imperfect, and that new symmetries or new dynamical principles are required to make it fully robust. Throughout the narrative, I emphasize concepts over technical details.

%
%
%

 \section{EXPERIMENTAL ROOTS OF THE ELECTROWEAK THEORY \label{sec:eroots}}
This section is devoted to a compressed evocation of how the phenomenology of the (charged-current) weak interactions developed, in order to establish what a successful theory would need to explain. A superb source for the experimental observations that led to the creation of the standard model is the book by  Cahn \& Goldhaber~\cite{CGExpFound}, which discusses and reproduces many classic papers.

Becquerel's discovery~\cite{Becquerel:1896zz}  of radioactivity in 1896 is one of the
wellsprings of modern physics.  In a short time, physicists learned 
to distinguish several sorts of radioactivity, classified by 
Rutherford~\cite{BaronRutherford} according to the character of the energetic projectile emitted in the spontaneous 
disintegration.  Natural and artificial radioactivity includes nuclear 
$\beta$ decay, observed as
\begin{equation}
	^{A}{\mathrm{Z}} \to\ ^{A}({\mathrm{Z+1}}) + \beta^{-}\; ,
	\label{eq:betadk}
\end{equation}
where $\beta^{-}$ is Rutherford's name for what was soon identified 
as the electron and
$^{A}{\mathrm{Z}}$ stands for the nucleus with charge $Z$ and mass number $A$ (in modern language, $Z$ protons and $A-Z$ neutrons).  Examples are tritium $\beta$ decay, $^{3}\mathrm{H}_{1} \to\ ^{3}\mathrm{He}_{2} + \beta^{-}$, neutron $\beta$ decay, $n \to p + \beta^{-}$, and $\beta$ decay of Lead-214, $^{214}\mathrm{Pb}_{82} \to\ ^{214}\mathrm{Bi}_{83} + \beta^{-}$.

For  two-body decays, as indicated by the detected products, the Principle of Conservation of Energy \&
Momentum says that the $\beta$ particle should have a definite energy. What was 
observed, as experiments matured, was very different: in 1914, James Chadwick~\cite{Chadwick:1914zz} (later to
discover the neutron) showed conclusively that in the
decay of Radium B and C ($^{214}$Pb and $^{214}$Bi), the 
$\beta$ energy follows a continuous spectrum.

The $\beta$-decay energy crisis tormented physicists for years.  On 
December 4, 1930, Wolfgang Pauli addressed an open letter~\cite{Paulinu} to a meeting on radioactivity in 
T\"{u}bingen.  In his letter, 
Pauli advanced the outlandish idea of a new, very penetrating, neutral 
particle of vanishingly small mass.  Because Pauli's new particle 
interacted very feebly with matter, it would escape undetected from 
any known apparatus, taking with it some energy, which would seemingly 
be lost.  The balance of energy and momentum would be restored by the
particle we now know as the electron's antineutrino.  Accordingly, the proper 
scheme for beta decay is 
\begin{equation}
	^{A}{\mathrm{Z}} \to\ ^{A}({\mathrm{Z+1}}) + \beta^{-} + \bar{\nu} \; .
	\label{eq:betadknu}
\end{equation}
What Pauli called his ``desperate remedy'' was, 
in its way, very conservative, for it preserved the principle of 
energy and momentum conservation and with it the notion that the laws 
of physics are invariant under translations in space and time.

After Chadwick's discovery of the neutron in
1932 in highly penetrating radiation emitted by beryllium irradiated by $\alpha$ particles~\cite{Chadwick:1932ma}, Fermi named Pauli's hypothetical particle the neutrino, to distinguish it
from Chadwick's strongly interacting neutron, and constructed his four-fermion theory (what we would today call a low-energy effective theory) of $\beta$ decay, which was the first step toward the modern theory of the charged-current weak interaction \cite{Fermi:1934sk}.  In retrospect, nuclear $\beta$ decay was the first hint for \textit{flavor,} the existence of particle families containing distinct species. That hint was made manifest by the discovery of the neutron, nearly degenerate in mass with the proton, which suggested that neutron and proton might be two  states
of a \textit{nucleon,} with the $n$ - $p$ mass difference attributed to electromagnetic effects. The inference that neutron and proton were partners was strengthened by the observation that nuclear forces are charge-independent, up to electromagnetic corrections~\cite{Quigg:2002td}. The accumulating evidence inspired Heisenberg~\cite{Heisenberg:1932dw} and Wigner~\cite{PhysRev.51.106} to make an analogy between the proton and neutron on the one hand and the up and down spin states of an electron. \textit{Isospin symmetry,} based on the spin-symmetry group \isogg, is the first example of a \textit{flavor symmetry.}

Detecting a
particle that interacts as feebly as the neutrino requires a massive target and a
copious source of neutrinos.  In 1953, Clyde Cowan and Fred Reines~\cite{Reines:1953pu} 
used the intense flux of antineutrinos from a fission
reactor and a heavy target ($10.7~\mathrm{ft}^{3}$ of liquid
scintillator) containing about $10^{28}$ protons to detect the inverse neutron-$\beta$-decay
reaction $\bar{\nu} + p \to\ e^{+} + n$.  Initial runs at the Hanford
Engineering Works were suggestive but inconclusive.  Moving their
apparatus to the stronger fission neutrino source at the Savannah
River nuclear plant, Cowan and Reines and their team made the
definitive observation of inverse $\beta$ decay in 1956~\cite{Cowan:1992xc}.

Through the 1950s, a series of experimental puzzles  led to the
suggestion that the weak interactions did not respect reflection
symmetry, or parity~\cite{Lee:1956qn}.  In 1956, C. S. Wu and collaborators detected a
correlation between the spin vector $\vec{J}$ of a polarized
$^{60}\mathrm{Co}$ nucleus and the direction $\hat{p}_{e}$ of the
outgoing $\beta$ particle~\cite{Wu:1957my}. Now, parity inversion
leaves spin, an axial vector, unchanged
($\mathcal{P}: \vec{J} \to \vec{J}$)
while reversing the electron direction
($\mathcal{P}: \hat{p}_{e} \to - \hat{p}_{e}$), so
the correlation $\vec{J} \cdot \hat{p}_{e}$ should be an ``unobservable'' null quantity if parity is a good symmetry. The observed correlation is \textit{parity 
violating.}  Detailed analysis of the $^{60}\mathrm{Co}$ result and 
others that came out in quick succession established that the 
charged-current weak interactions are left-handed.  By the same argument, the parity operation
links a left-handed neutrino with a right-handed neutrino. Therefore a theory that contains only $\nu_{\mathrm{L}}$ would be manifestly parity-violating.

Could the neutrino indeed be left-handed? M.~Goldhaber and collaborators inferred the electron neutrino's helicity~\cite{Goldhaber:1958nb} from the
longitudinal polarization of the recoil nucleus in the
electron-capture reaction
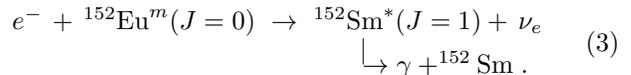
\dknuc{e^{-}\, +\, ^{152}\mathrm{Eu}^{m} (J=0)}{^{152}\mathrm{Sm}^{*} 
(J=1)+\,\nu_{e}}{\gamma + ^{152}\mathrm{Sm}\; .}
A compendious knowledge of the properties of nuclear levels, together with meticulous technique, enabled this classic experiment. 

Following the observation of maximal parity violation in the late 1950s, a 
serviceable effective Lagrangian for the weak interactions of 
electrons and neutrinos could be written as the product of charged 
leptonic currents,
\begin{equation}
    \lag_{\mathrm{V-A}} = \frac{-\gf}{\sqrt{2}}\bar{\nu}\gamma_{\mu}
    (1 - \gamma_{5})e\; \bar{e}\gamma^{\mu}(1 - \gamma_{5})\nu
    + \mathrm{h.c.}\; ,
    \label{eq:lepel}
\end{equation}
where Fermi's coupling constant is $\gf = 1.166 3787(6) \times 
10^{-5}\gev^{-2}$.  This Lagrangian has a $\mathrm{V-A}$ (vector minus axial 
vector) Lorentz structure~\cite{Feynman:1958ty,Gershtein:1955fb,Sudarshan:1958vf,Sakurai:1958zz}, whereas Fermi's effective Lagrangian for $\beta$ decay was a (parity-conserving) vector interaction. A straightforward lepton-current--times--nucleon-current generalization of \Eqn{eq:lepel} that takes account of the fact that  nucleons are not simple Dirac particles leads to an effective Lagrangian for $\beta$ decay and associated processes. Many applications and experimental tests are detailed in~\cite{Marshak,Commins,ComminsBucksbaum}.

The direct phenomenological consequences of parity violation in the weak interactions, which shattered the received wisdom of the era, were themselves dramatic, leading for example to a factor-of-three difference between the total cross sections for $\nu e$ and $\bar{\nu} e$ scattering. Parity violation is also a harbinger of a particular challenge to be met by a true theory of the weak interactions. In quantum electrodynamics, it is perfectly respectable (and correct!) to write a Lagrangian
that includes a term for electron mass,
\begin{equation}
	\lag = \bar{e}(i\gamma^{\mu}\D_\mu - m)e = 
	  \bar{e}(i\gamma^{\mu}\partial_\mu - m)e - 
	 qA_\mu\bar{e}\gamma^\mu e ,
	\label{lagQED}
\end{equation} where $A_\mu$ is the four-vector potential of electromagnetism.
The left-handed and right-handed components of the electron have the same charge, and so appear symmetrically. If fermions are \textit{chiral,} which is to say that the left-handed and right-handed components behave differently, a mass term conflicts with symmetries. This will be made precise in  \S\ref{GlashowTh}.

A second charged lepton, the muon, was discovered and identified as lacking strong interactions in the decade beginning in 1937~\cite{Neddermeyer:1937md,Street:1937me,Conversi:1947ig}. In common with the electron, the muon is a spin-$\frac{1}{2}$ Dirac particle, structureless at our present limits of resolution. It is unstable, with a mean lifetime of approximately $2.2\;\mu\mathrm{s}$ and a mass $207 \times$ that of the electron. It might be tempting, therefore, to consider the muon as an excited electron, but the transitions $\mu \to e \gamma$, $\mu \to e e^+e^-$, and $\mu \to e \gamma\gamma$ has never been seen. The limits on these decays are so stringent [e.g., the branching fraction for $\mu \to e \gamma$ is $< 5 \times 10^{-13}$ at 90\% confidence level~\cite{Agashe:2014kda}] that we regard the muon as a distinct lepton species.

If the muon is distinct from the electron, what is the nature of the 
neutrino produced in association with the muon in pion decay, $\pi^{+} \to \mu^{+}\nu$?  In 1962, 
Lederman, Schwartz, Steinberger, and collaborators  carried out a 
\textit{two-neutrino experiment} using neutrinos created in the decay 
of high-energy pions from the new Alternating Gradient Synchrotron at 
Brookhaven~\cite{Danby:1962nd}.  They observed numerous examples of the 
reaction $\nu N \to \mu + X$, but found no evidence for the 
production of electrons.  Their study established that the muon 
produced in pion decay is a distinct particle, $\nu_{\mu}$, that is 
different from either $\nu_{e}$ or $\bar{\nu}_{e}$.  This observation 
suggests that the leptonic charged-current weak interactions exhibit 
 a two-doublet family structure,
\begin{equation}
 			\left(
		\begin{array}{c}
			\nu_{e}  \\
			e^{-}
		\end{array}
		 \right)_{\mathrm{L}}\qquad
		\left(
		\begin{array}{c}
			\nu_{\mu}  \\
			\mu^{-}
		\end{array}
		 \right)_{\mathrm{L}} \; .
    \label{eq:famlep}
\end{equation}
We are led to generalize the effective Lagrangian \eqn{eq:lepel} to 
include the terms
\begin{equation}
    \lag_{\mathrm{V-A}}^{(e\mu)} = \frac{-\gf}{\sqrt{2}}\bar{\nu}_{\mu}\gamma_{\mu}
    (1 - \gamma_{5})\mu\; \bar{e}\gamma^{\mu}(1 - \gamma_{5})\nu_{e}
    + \mathrm{h.c.}\; ,
    \label{eq:lepmueel}
\end{equation}
in the familiar current-current form.

Because the weak interaction acts at a point, the effective Lagrangians hold only over a finite range of energies, and cannot reliably be computed beyond leading order. A classic application~\cite{Lee:1965js} of partial-wave unitarity (probability conservation) to inverse muon decay, $\nu_\mu e \to \mu \nu_e$, leads to the conclusion that the four-fermion effective Lagrangian \Eqn{eq:lepmueel} can only make sense for c.m.\ energies $\sqrt{s} \le 617\gev$. That comfortably encompasses most laboratory experiments, but as a matter of principle gives a clear lesson: new physics must intervene below about $600\gev$ c.m.\ energy.

Although Fermi took his inspiration from the theory of electromagnetism, he did not posit a force carrier analogous to the photon. This is a perfectly reasonable first step, given that electromagnetism acts over an infinite range, whereas the influence of the $\beta$-decay interaction extends only over about $10^{-15}\cm.$ One may hope to obtain a more satisfactory theory by taking the next step,  supposing that the weak interaction, like quantum electrodynamics, is mediated by vector-boson exchange (of nonzero range) to soften the high-energy growth of amplitudes. The weak intermediate boson must carry charge $\pm 1$, because the familiar manifestations of the weak interactions (such as $\beta$-decay) are charge-changing; be rather massive $(\approx 100\gev)$, to reproduce the short range of the weak force; and accommodate parity violation. Introducing a weak boson $W^\pm$ in this \textit{ad hoc} manner indeed mitigates the unitarity problem for inverse muon decay, but introduces incurable unitarity problems for reactions such as $e^+ e^- \to W^+ W^-$ or $\nu \bar{\nu} \to W^+ W^-$, as detailed in \S6.2 of~\cite{GT2}.

It is also worth mentioning the discovery of strange particles in the early 1950s, because it was essential to establishing that the leptonic and hadronic weak interactions have the same strength and stimulated the invention of quarks~\cite{Zweig:1981pd,Zweig:1964jf,Gell-Mann:1964nj}. Semileptonic decays of hyperons~\cite{Cabibbo:2003cu} were an essential testing ground for Cabibbo's formulation of the universality of the charged-current weak interactions~\cite{Cabibbo:1963yz} which was the forerunner of today's $3 \times 3$ quark-mixing matrix~\cite{Kobayashi:1973fv}.

\section{THE DEVELOPMENT OF THE ELECTROWEAK THEORY \label{sec:EWTheory}}
This section is a brief historical survey of the ideas that came together in the notion of a gauge theory for the weak and electromagnetic interactions. What follows is neither a complete intellectual history (which would occupy a book) nor an abbreviated course, but a tour of key themes, including Yang--Mills theory, the insight from superconductivity that spontaneous breaking of a gauge symmetry endows gauge bosons with mass, and the development of the electroweak theory as we know it. The aim here is to stress the interplay of ideas from diverse sources and to show how the electroweak theory responds to the established phenomenology of the weak interactions.

 \subsection{Symmetries and Interactions \label{subsec:SymInt}}
 Notions of symmetry lie at the heart of much of science, and a confidence in the importance of symmetry is a guiding principle for scientists in many disciplines. Werner Heisenberg's quasi-Biblical pronouncement, \guillemotright Am Anfang war die Symmetrie\guillemotleft~[``In the beginning was Symmetry'']~\cite{WernerH}, resonates in much theoretical work from the early twentieth century to the present. An essential insight of our modern conception of nature is that \textit{symmetries dictate interactions.} 
 
While Heisenberg's assertion can be challenged as mere opinion, physicists have learned over the past century how to connect symmetries with conservation laws, and symmetries with interactions.   The 1918 work of Emmy Noether~\cite{EmmyN,NinaB} which took the form of two mathematical theorems, showed that to every continuous global symmetry of the laws of nature there corresponds a conservation law. Thus, translation invariance in space---the statement that the laws are the same everywhere---is connected with conservation of momentum. Invariance under translations in time is correlated with the conservation of energy. Invariance under rotations implies the conservation of angular momentum. Noether's theorem shows how conservation laws could arise, and indeed how they could be exact statements, not merely summaries of empirical evidence.

 The derivation of interactions from symmetries was initiated by Weyl~\cite{HermannW} in a series of papers from 1918 to 1929, spanning the invention of quantum mechanics. In the version that became a prototype for modern gauge theories, Weyl showed that by requiring that the laws of nature be invariant under local changes of the phase convention for the quantum-mechanical wave function, $\psi(x) \rightarrow \psi^\prime(x) = e^{i\alpha(x)} \psi(x)$, one could derive the laws of electrodynamics. Invariance under global (coordinate-independent) \emgg\ phase rotations implies the conservation of electric charge; invariance under local (coordinate-dependent) \emgg\ phase rotations implies the existence of a massless vector field---the photon---that couples minimally to the conserved current of the theory. A straightforward derivation leads to the Lagrangian,
 \begin{eqnarray}
 \lag_{\mathrm{QED}} & = & \lag_{\mathrm{free}} - J^\mu A_\mu - {\textstyle\frac{1}{4}} 
	F_{\mu\nu}F^{\mu\nu} \label{lagQED2}\\
	 & = & \bar{\psi}(i\gamma^{\mu}\partial_\mu - m)\psi - 
	 qA_\mu\bar{\psi}\gamma^\mu\psi - {\textstyle\frac{1}{4}} 
	F_{\mu\nu}F^{\mu\nu}\nonumber ,  
\end{eqnarray} 
where $\psi$ is the electron field, $\partial_\mu + iqA_\mu(x) \equiv \D_\mu$  is the gauge-covariant derivative, $J^\mu = q\bar{\psi}\gamma^\mu\psi$ is the conserved electromagnetic current, and the field-strength tensor is $F^{\mu\nu} = -F^{\nu\mu} = \partial^\nu A^\mu - \partial^\mu A^\nu$.  The $F_{\mu\nu}F^{\mu\nu}$ term, which accounts for photon propagation, is called the kinetic term. Under a local phase rotation, the photon field transforms as $A_\mu(x) \rightarrow A_\mu(x) - \partial_\mu \alpha(x)$, the familiar form of a gauge transformation in (even classical) electrodynamics. The electron mass term ($-m\bar{\psi}\psi$) respects the local gauge symmetry. A photon mass term would have the form $\lag_\gamma = \frac{1}{2} m^2 A^\mu A_\mu$,
 which conflicts with local gauge invariance because $A^\mu A_\mu \rightarrow 
	(A^\mu - \partial^\mu \alpha)(A_\mu - \partial_\mu\alpha)
	\ne A^\mu A_\mu$.  Thus has local gauge invariance  led to the 
existence of a massless photon.

 The construction of Quantum Electrodynamics as the gauge theory~\cite{Jackson:2001ia} based on \emgg\ phase symmetry provides a template for building other interactions derived from symmetries. In 1954, as isospin emerged as a reliable classification symmetry for nuclear levels and as a tool for understanding nuclear forces, Yang \& Mills~\cite{Yang:1954ek} [see also Shaw~\cite{ShawYM}] asked whether isospin, promoted to a local symmetry, could lead to a theory of nuclear forces. It is a lovely idea: derive the strong interactions among nucleons by requiring that the theory be invariant under independent choices at every point of the convention defining proton and neutron.
 
 The construction begins with the free-nucleon Lagrangian
\begin{equation}		\lag_{0} = \bar{\psi}(i\gamma^\mu\partial_\mu - m)\psi  ,
\label{eq:freenucleon}
\end{equation} 
	 written in terms of the composite fermion fields
$	\psi \equiv {\left(
	\begin{smallmatrix}
		p  \\
		n
	\end{smallmatrix}
	\right)} .
	 $  The Lagrangian 
\eqn{eq:freenucleon} is invariant under global isospin rotations $\psi 
\to 
\exp{\left(i\mbox{{\boldmath$\tau$}$\cdot${\boldmath$\alpha$}/2}\right)}
\psi$, where  \mbox{\boldmath$\tau$} is a Pauli isospin matrix, and the 
isospin current ${\mathbf{J}}^\mu = \bar{\psi}\gamma^\mu 
	\frac{{\mbox{\boldmath $\tau$}}}{2} \psi$ is conserved.  Now require invariance under a local gauge transformation, $\psi(x) \rightarrow \psi^\prime(x) = G(x)\psi(x)$, with 
	$G(x) \equiv \exp\left(i \btau\cdot\balpha(x)/2\right)$. The construction is similar to the one made for QED, but is more involved because of the non-Abelian nature of the \isogg\ isospin gauge group. In this case, we find an isovector of gauge fields, corresponding to the adjoint representation of \isogg. The gauge fields satisfy the transformation law $\mathbf{b}_\mu^\prime = \mathbf{b}_\mu - \balpha\times \mathbf{b}_\mu -  (1/g)\partial_\mu\balpha$ or, in component form, $b_\mu^{\prime\,l} = b_\mu^l - \varepsilon_{jkl}\alpha^j b^k - (1/g)\partial_\mu\alpha^l$: the translation familiar from QED plus an isospin rotation. Here $g$ is the coupling constant of the theory. The field-strength tensor is $F_{\mu\nu}^l = \partial_\nu b_\mu^l - \partial_\mu b_\nu^l
	+ g\varepsilon_{jkl}b_\mu^j b_\nu^k$. It is convenient to define $F_{\mu\nu} = \frac{1}{2} 
	F_{\mu\nu}^l\tau^l $. Then we may write the Yang--Mills Lagrangian as
\begin{eqnarray}
	\lag_{\mathrm{YM}} & = & \bar{\psi}(i\gamma^\mu\partial_\mu -m)\psi -\frac{g}{2}\mathbf{b}_\mu\cdot 
	 \bar{\psi}\gamma^\mu\btau\psi - {\textstyle\frac{1}{2}}\tr{F_{\mu\nu}F^{\mu\nu}}
	\nonumber  \\
	 & = & \lag_0 -\frac{g}{2}\mathbf{b}_\mu\cdot 
	 \bar{\psi}\gamma^\mu\btau\psi - {\textstyle\frac{1}{2}}\tr{F_{\mu\nu}F^{\mu\nu}}, 
	\label{ymnucL}
\end{eqnarray} namely a free Dirac Lagrangian plus an interaction term 
that couples the isovector gauge fields to the conserved isospin current, plus a kinetic term that now describes both the propagation and the self-interactions of the gauge fields. As in the case of electromagnetism, a mass term quadratic in the gauge fields is incompatible with local gauge 
invariance, as in electromagnetism, but nothing forbids a common nonzero mass for the 
nucleons. The quadratic term in the gauge fields present in the field-strength tensor gives rises to self-interactions among the gauge bosons that are not present in Abelian theories such as QED.

The discovery that interactions may be derived from isospin symmetry, and from a general gauge group~\cite{Glashow:1961ep}, provides theorists with an important strategy for deriving potentially well-behaved theories of the fundamental interactions. Nuclear forces are not mediated by massless spin-1 particles, so the Yang--Mills theory does not succeed in the goal that motivated it. Nevertheless, the approach underlies two new laws of nature: quantum chromodynamics and the electroweak theory.
\subsection{\ewgg \label{GlashowTh}}
The Yang--Mills experience shows that there is no guarantee that a gauge theory built on a particular symmetry will faithfully describe some aspect of matter. A great deal of art and, to be sure, trial and error, goes into the selection of the right gauge symmetry. In the late 1950s and early 1960s, several authors advanced proposals for a gauge theory of the weak interactions, or of a ``unified'' theory of the weak and electromagnetic interactions, reading clues from experiment as best they could. Even after what would become the standard \ewgg\ electroweak theory had emerged and was elaborated, imaginative theorists put forward alternative ideas, guided either by experimental hints or by aesthetics. We do not (yet) have a way of deducing the correct gauge symmetry from higher principles.

What turned out to be the correct choice was elaborated by Glashow in 1961~\cite{Glashow:1961tr}. Let us review the essential structure to recall why a new idea was needed to arrive at a successful theory, even after the correct symmetry had been chosen. The leptonic elements of the theory will suffice to exhibit the motivation and the principal features.

We begin by designating the spectrum of fundamental fermions of the theory. It suffices for the moment to include only the electron and its neutrino, which form a left-handed ``weak-isospin'' doublet (cf. \Eqn{eq:famlep}),
$\mathsf{L}_e \equiv {\left( \begin{smallmatrix} \nu \\ e \end{smallmatrix} \right)_{\mathrm{L}}}$,
where the left-handed states are $\nu_{\mathrm{L}} = \frac{1}{2}(1 - \gamma_5)\nu$ and $e_{\mathrm{L}} = \frac{1}{2}(1 - \gamma_5)e$.  For the reasons we reviewed in \S\ref{sec:eroots}, it is convenient to assume that the right-handed state $\nu_{\mathrm{R}} = \frac{1}{2}(1 + \gamma_5)\nu$
does not exist. Thus we designate only one right-handed lepton, $\mathsf{R}_e = e_{\mathrm{R}} = \frac{1}{2}(1 + \gamma_5)e$,
which is a weak-isospin singlet. This completes a specification of the charged weak currents.

To incorporate electromagnetism, Glashow defines a ``weak hypercharge,'' $Y$. Requiring that the Gell-Mann--Nishijima relation for the electric charge, $Q = I_3 + \frac{1}{2} Y$,
be satisfied leads to the assignments
$Y_{\mathrm{L}} = -1$ and $Y_{\mathrm{R}} =-2$. By construction, the weak-isospin projection $I_3$ and the weak hypercharge $Y$ are commuting observables.

We now take the (product) group of transformations generated by $I$ and $Y$ to be the gauge group \ewgg\ of the theory. To construct the theory, we introduce the gauge fields
\begin{equation}
\begin{array}{rcl}
b_\mu^1, b_\mu^2, b_\mu^3 & \phantom{MMM} & \hbox{for }\wigg , \\
\mathcal{A}_\mu & \phantom{MMM} & \hbox{for }\ygg .
\end{array}
\label{eq:lepgbs}
\end{equation}
Evidently the Lagrangian for the theory may be written as
\begin{equation}
\lag = \lag_{\mathrm{gauge}} + \lag_{\mathrm{leptons}}
\label{eq:ewlag}
\end{equation}
where the kinetic term for the gauge fields is
\begin{equation}
 \lag_{\mathrm{gauge}} = -{\textstyle\frac{1}{4}}F^l_{\mu\nu}F^{l\mu\nu} - {\textstyle\frac{1}{4}} f_{\mu\nu}f^{\mu\nu}
 \label{eq:ewglag}
 \end{equation}
and the field-strength tensors are $F^l_{\mu\nu} = \partial_\nu b^l_\mu - \partial_\mu b^l_\nu + g\varepsilon_{jkl}b^j_\mu b^k_\nu$ for the \wigg\ gauge fields and
$f_{\mu\nu} = \partial_\nu\mathcal{A}_\mu - \partial_\mu\mathcal{A}_\nu$ for the \ygg\ gauge field.  The matter term is
\begin{eqnarray}
\lag_{\mathrm{leptons}} & = & \bar{\mathsf{R}}_ei\gamma^\mu\left( \partial_\mu + \frac{ig^\prime}{2}\mathcal{A}_\mu Y\right)\mathsf{R}_e \\ &  + & \bar{\mathsf{L}}_ei\gamma^\mu\left( \partial_\mu + \frac{ig^\prime}{2}\mathcal{A}_\mu Y + \frac{ig}{2}\dotp{\btau}{\mathbf{b}_\mu}\right)\mathsf{L}_e \nonumber.
\label{eq:leplag}
\end{eqnarray}
The coupling of the weak-isospin group \wigg\ is called $g$, as in the Yang-Mills theory,  and the coupling constant for the weak-hypercharge group \ygg\ is denoted as $g^\prime/2$. the factor $\frac{1}{2}$ being chosen to simplify later expressions. Similar structures appear for the hadronic weak interactions, now expressed in terms of \textit{quarks.} The universal strength of charged-current interactions follows from the fact that both the left-handed quarks and the left-handed leptons reside in weak-isospin doublets.

The theory of weak and electromagnetic interactions described by the Lagrangian \Eqn{eq:ewglag} is not a satisfactory one, for two immediately obvious reasons. It contains four massless gauge bosons $(b^1, b^2, b^3, \mathcal{A})$, whereas Nature has but one, the photon. In addition, the expression \eqn{eq:leplag} represents a massless electron; it lacks the $-m_e \bar{e}e$ term of the QED Lagrangian \Eqn{lagQED2}, and for good reason. A fermion mass term links left-handed and right-handed components: $\bar{e}e = \frac{1}{2}\bar{e}(1 - \gamma_5)e + \frac{1}{2}\bar{e}(1 + \gamma_5)e = \bar{e}_{\mathrm{R}}e_{\mathrm{L}} + \bar{e}_{\mathrm{L}}e_{\mathrm{R}}$.
The left-handed and right-handed components of the electron transform differently under \wigg\ and \ygg, so an explicit fermion mass term would break the \ewgg\ gauge invariance of the theory: such a mass term is forbidden. 

\subsection{Insights from Superconductivity \label{subsec:Supercon}}  
How gauge bosons can acquire mass is a conundrum for both  the Yang--Mills theory as a description of nuclear forces and for the \ewgg\ theory as a description of the weak and electromagnetic interactions. An important general insight is that the symmetries of the laws of nature need not be manifest in the outcome of those laws. Hidden (or secret) symmetries are all around us in the everyday world---for example, in the ordered structures of crystals and snowflakes or the spontaneous magnetization of a ferromagnetic substance, configurations that belie the $\mathrm{O(3)}$ rotation symmetry of electromagnetism. [See~\cite{AnderGauge} for an interesting tour of spontaneous symmetry breaking in many physical contexts.] The common feature of these phenomena is that the symmetry exhibited by the state of lowest energy, the vacuum, is not the full symmetry of the theory. In addition, the vacuum is degenerate, characterized by many states of the same energy, and the choice of any one is aleatory.

As it happens, \textit{superconductivity,} a rich and fascinating phenomenon from condensed-matter physics, points the way to understanding how gauge bosons can acquire mass. In 1911, shortly after he succeeded in liquefying helium, and therefore could conduct experiments at unprecedented low temperatures, Heike Kamerlingh Onnes~\cite{HKO} observed the sudden vanishing of electrical resistance in a sample of mercury cooled to 4.2~K. This first miracle of superconductivity is of immense technological importance, not least in magnets that are essential components of the Large Hadron Collider.

The second miracle, which for me marks superconductivity as truly extraordinary, was discovered in 1933 by Meissner \& Ochsenfeld~\cite{Meissner} [for an English translation, see~\cite{0143-0807-4-2-011}]: magnetic flux is excluded from the superconducting medium. A typical penetration depth~\cite{London2} is on the order of $10~\mu\mathrm{m}$. This means that, within the superconductor, \textit{the photon has acquired a mass.} Here is the germ of the idea that leads to understanding how the force particles in gauge theories could be massive: QED is a gauge theory, and under the special circumstances of a superconductor, the normally massless photon becomes massive, while electric charge remains a conserved quantity.

Two decades would pass before the idea would be fully formed and ready for application to theories of the fundamental interactions. The necessary developments included the elaboration of relativistic quantum field theory and the full realization of QED, a focus on the consequences of spontaneous symmetry breaking, the emergence of informative theories of superconductivity, and attention to the special features of gauge theories.

\subsection{Spontaneous Symmetry Breaking \label{subsec:ssb}}
A key insight into hidden symmetry in field theory was achieved by Jeffrey Goldstone~\cite{Goldstone:1961eq}, who considered the Lagrangian for two scalar fields $\phi_1$ and $\phi_2$,
\begin{equation}
	\lag = {\textstyle\frac{1}{2}}[(\partial_\mu\phi_1)(\partial^\mu\phi_1) +
	(\partial_\mu\phi_2)(\partial^\mu\phi_2)] -V(\phi_1^2+\phi_2^2).
	\label{Lgold}
\end{equation}  The Lagrangian is invariant under the group $\mathrm{SO(2)}$ of 
rotations in the $\phi_1$-$\phi_2$ plane.
It is informative to consider the effective potential
\begin{equation}
	V(\bphi^2) = {\textstyle\frac{1}{2}}\mu^2\bphi^2 + 
	{\textstyle\frac{1}{4}}\abs{\lambda} (\bphi^2)^2,
	\label{goldpot}
\end{equation} where $\bphi = \left(
\begin{smallmatrix}
\phi_1 \\
\phi_2
\end{smallmatrix}
\right)$ and $\bphi^2=\phi_1^2+\phi_2^2$, and 
distinguish two cases.  

A positive value of the parameter $\mu^2>0$ corresponds to the ordinary 
case of unbroken symmetry.  The unique minimum, corresponding to the vacuum 
state, occurs at	$\vev{\bphi} = \left(
	\begin{smallmatrix}
		0  \\		
		0
	\end{smallmatrix}
	\right)$, and so for small oscillations the Lagrangian takes the form
\begin{equation}
	\lag_{\mathrm{so}} = {\textstyle\frac{1}{2}}[(\partial_\mu\phi_1)(\partial^\mu\phi_1) 
	-\mu^2\phi_1^2] + {\textstyle\frac{1}{2}}[(\partial_\mu\phi_2)(\partial^\mu\phi_2) 
	-\mu^2\phi_2^2] ,
	\label{goldSO}
\end{equation}  which describes
a pair of scalar particles with common mass $\mu$.  Thus the introduction 
of a symmetric interaction preserves the spectrum 
of the free theory with $\abs{\lambda}=0$. 

For the choice $\mu^2 < 0$, a line of minima lie along $\vev{\bphi^2} = -\mu^2/\abs{\lambda} \equiv v^2$, a continuum of distinct vacuum 
states, degenerate in energy.  The degeneracy follows from the $\mathrm{SO(2)}$ symmetry of the potential (\ref{Lgold}).  
Designating one state as the vacuum selects a preferred direction in 
$(\phi_{1},\phi_{2})$ internal symmetry space, and amounts to a spontaneous 
breakdown of the $\mathrm{SO(2)}$   
symmetry. Let us select as the 
physical vacuum state the configuration $\vev{\bphi} = \left(
	\begin{smallmatrix}
		v  \\
		0
	\end{smallmatrix}
	\right)$,
 as we may always do with a suitable definition of 
coordinates.  Expanding about the vacuum configuration by defining $\bphi^\prime \equiv \bphi - 
	\vev{\bphi} \equiv \left(
	\begin{smallmatrix}
		\eta \\		
		\zeta  		
	\end{smallmatrix}
	\right)$, we obtain the Lagrangian for small oscillations
\begin{equation}
	\lag_{\mathrm{so}} = {\textstyle\frac{1}{2}}[(\partial_\mu\eta)(\partial^\mu\eta) + 
	2\mu^2\eta^2] + {\textstyle\frac{1}{2}}[(\partial_\mu\zeta)(\partial^\mu\zeta)],
	\label{goldSBSO}
\end{equation} plus an irrelevant constant.  There are still two particles in 
the spectrum.  The $\eta$-particle, associated with radial oscillations, 
has $(\mathrm{mass})^2 = -2\mu^2 > 0$.  The 
$\zeta$-particle, however, is massless.  The mass of the $\eta$-particle 
may be viewed as a consequence of the restoring force of the potential 
against radial oscillations.  In contrast, the masslessness of the 
$\zeta$-particle is a consequence of the $\mathrm{SO(2)}$ invariance of the 
Lagrangian, which means that there is no restoring force against angular 
oscillations. It is ironic that the $\eta$-particle, which here seems so unremarkable, is precisely what emerges as the ``Higgs boson'' when the hidden symmetry is a gauge symmetry.

The splitting of the spectrum and the appearance of the 
massless particle are known as the Goldstone phenomenon.  Such massless 
particles, zero-energy excitations that connect possible vacua, are called Nambu--Goldstone bosons. Many occurrences are known in particle, nuclear, and condensed-matter physics~\cite{Burgess:1998ku}. In any field theory that
obeys the ``usual axioms,'' including locality, Lorentz invariance, and
positive-definite norm on the Hilbert space, if an exact continuous
symmetry of the Lagrangian is not a symmetry of the physical vacuum,
then the theory must contain a massless spin-zero particle (or
particles) whose quantum numbers are those of the broken group
generator (or generators)~\cite{Goldstone:1962es}.

This strong statement seemed a powerful impediment to the use of spontaneous symmetry breaking in realistic theories of the fundamental interactions, as the disease of unobserved massless spin-0 particles was added to the disease of massless gauge bosons. Motivated by analogy with the plasmon
theory of the free-electron gas  Anderson~\cite{Anderson:1963pc} put forward a prescient conjecture that one zero-mass ill might cancel the other and make possible a realistic Yang--Mills theory of the strong interactions.

The decisive contributions came at a time of intense interest in superconductivity---in the intricacies of the Bardeen--Cooper--Schrieffer (BCS) theory~\cite{Bardeen:1957mv} and in understanding the role of symmetry breaking in the Meissner effect. From the remove of a half century, it seems to me that  preoccupation with the microscopic BCS theory might have complicated the search for a cure for the massless gauge bosons. An easier path is to analyze the phenomenological Ginzburg--Landau~\cite{Ginzburg:1950sr,VLG} description of the superconducting phase transition in the framework of QED. It is then easy to see how the photon acquires mass in a superconducting medium [see, e.g., Problem 5.7 of Ref. ~\onlinecite{GT2}, \S21.6 of Ref. ~\onlinecite{WeinbergQFT} and the ``Abelian Higgs Model,''~\cite{Higgs:1966ev}]. But that is hindsight and speculation!

Searching for a solution to the problem of massless gauge bosons in field theory, Englert \& Brout~\cite{Englert:1964et}, Higgs~\cite{Higgs:1964ia,Higgs:1964pj}, and Guralnik, Hagen, \& Kibble~\cite{Guralnik:1964eu} showed that gauge theories are different. They do not 
satisfy the assumptions on which Goldstone theorem is based, although they are 
respectable field theories.  Recall that to quantize electrodynamics, an exemplary gauge theory,
one must choose between the covariant Gupta--Bleuler formalism with its 
unphysical indefinite-metric states or quantization in a physical gauge 
for which manifest covariance is lost. Through their work, we understand that the would-be Goldstone bosons that correspond to broken generators of a gauge symmetry become the longitudinal components of the corresponding gauge bosons. What remains as scalar degrees of freedom is an incomplete multiplet---defined by the unbroken generators of the gauge symmetry---of massive particles that we call Higgs bosons.

Their collective insight did not, as many had hoped, make a proper description of the strong nuclear force out of Yang--Mills theory. It did, however, set the stage for the development of the electroweak theory and for plausible, if still speculative, unified theories of the strong, weak, and electromagnetic interactions.

It is inaccurate to say that the work of these theorists solved a problem in the standard model---the standard model did not yet exist! Indeed, they were not concerned with the weak interactions, and the implications for fermion mass shifts are mentioned only in passing. [Recall that for nonchiral theories such as QED and the Yang-Mills theory, the origin of fermion masses does not arise, in the sense that fermion mass is consistent with the gauge symmetry.] Their work can be said to have triggered the conception of the electroweak theory, which is a very considerable achievement.

Following the discovery of the Higgs boson of the electroweak theory, Fran\c{c}ois Englert~\cite{RevModPhys.86.843} and Peter Higgs~\cite{RevModPhys.86.851} shared the 2014 Nobel Prize for Physics. Gerald Guralnik and Richard Hagen~\cite{Guralnik:2014bwa} have published a memoir of their work. In addition, several of the leading actors in the discovery of spontaneous gauge symmetry breaking as an origin of particle mass have described their personal involvement: Anderson~\cite{AndersonInterview},  Englert~\cite{Englert:2004yk}, Guralnik~\cite{doi:10.1142/S0217751X09045431,Guralnik:2011zz}, and Higgs~\cite{10.1063/1.45425,doi:10.1142/S0217751X02013046}. Their words carry a special fascination.
 
\subsection{The Electroweak Theory and the Standard-Model Higgs Boson \label{subsec:EWT}}
In the late 1960s, Steven Weinberg~\cite{Weinberg:1967tq} and Abdus Salam~\cite{Salam} used the new insights about spontaneous breaking of gauge symmetry to complete the program set out by Sheldon Glashow~\cite{Glashow:1961tr} that we recalled in \S\ref{GlashowTh}. The construction of the spontaneously broken \ewgg\ theory of the weak and electromagnetic interactions is detailed in many places, including \S2 of ``Unanswered Questions,'' Ref. ~\onlinecite{Quigg:2009vq}, so we may focus here on a few important conceptual matters.

If \ewgg\ proves to be the apt choice of gauge symmetry for a theory of weak and electromagnetic ineractions, then that symmetry must be hidden, or broken down to the \emgg\ symmetry we observe manifestly. The simple choice made by Weinberg and Salam, which now has significant empirical support, is to introduce a complex weak-isospin doublet of auxiliary scalar fields, and to contrive their self-interactions to create a degenerate vacuum that does not exhibit the full \ewgg\ symmetry. Before spontaneous symmetry breaking, we count eight degrees of freedom among the four massless gauge bosons and four degrees of freedom for the scalar fields. After spontaneous breaking of $\ewgg \to \emgg$, following the path we have just reviewed in \S\ref{subsec:ssb}, the scalar field obtains a vacuum-expectation value, $\vev{\phi} = \left(\begin{smallmatrix} 0 \\ v/\sqrt{2} \end{smallmatrix}\right)$, where $v = (\gf\sqrt{2})^{-\half} \approx 246\gev$ to reproduce the low-energy phenomenology. 

The charged gauge bosons, $W^\pm$, which mediate the $\mathrm{V-A}$ charged-current interaction, acquire mass $gv/2$. The neutral gauge bosons of \Eqn{eq:lepgbs} mix to yield a massive ($M_z = M_W/\cos\theta_{\mathrm{W}}$) neutral gauge boson, $Z^0$, that mediates a hitherto unknown weak-neutral-current interaction and a massless photon, $\gamma$. [The weak mixing angle $\theta_{\mathrm{W}}$, which parametrizes the mixing of $b_\mu^3$ and $\mathcal{A}_\mu$, is determined from experiment.] The photon has pure vector couplings, as required, whereas $Z^0$ has a mix of vector and axial-vector couplings that depend on the quantum numbers of the fermion in question. Eleven of the twelve bosonic degrees of freedom now reside in the vector bosons: $3\times3$ massive bosons + $1\times2$ massless photon. The last degree of freedom corresponds to the Higgs boson: it is a massive scalar, but the Weinberg--Salam theory does not fix its mass. Because the Higgs boson and the longitudinal components of the gauge bosons share a common origin, the Higgs boson plays an essential role in ensuring a sensible high-energy behavior of the electroweak theory~\cite{Lee:1977eg}.

All this is fixed by the construction of the theory: once the representation of the auxiliary scalar fields is chosen and the weak mixing parameter determined, all the couplings of gauge bosons to fermions and couplings among gauge and Higgs bosons are set. The new neutral-current interactions among the leptons are flavor-diagonal. Also, to this point, we have solved only one of the outstanding problems of the unbroken \ewgg\ theory: the masslessness of all the gauge bosons. What of the fermions? Weinberg and Salam saw the possibility to generate fermion masses in the spontaneously broken theory by adding to the Lagrangian a gauge-invariant  interaction, $\lag_{\mathrm{Yukawa}} = -\zeta_e[\bar{\mathsf{R}}_e(\phi^\dagger \mathsf{L}_e) + (\bar{\mathsf{L}}_e\phi)\mathsf{R}_e]$, where the Yukawa coupling, $\zeta_e$, is a phenomenological parameter. When the gauge symmetry is hidden, the Yukawa term becomes
\begin{equation}
\lag_{\mathrm{Yukawa}}  = - \frac{\zeta_e v}{\sqrt{2}}\,\bar{e}e - \frac{\zeta_e H}{\sqrt{2}}\,\bar{e}e ,
 \label{eq:lepyukbr}
 \end{equation}
where $H$ is the Higgs boson. The electron has acquired a mass $m_e = \zeta_e v/\sqrt{2}$, and the $He\bar{e}$ coupling is $-im_e/v$. It is pleasing that the electron mass arises spontaneously, but frustrating that the parameter $\zeta_e$ must be put in by hand, and does not emerge from the theory. The same strategy carries over for all the quarks and charged leptons, and may also be seen as the origin for the parameters of the quark-mixing matrix.

If the Higgs field is the source of the quark and charged-lepton masses, that does not mean that the ``Higgs boson is the source of all mass in the Universe,'' as is frequently stated---even by physicists. The overwhelming majority of the visible mass in the Universe is in the form of atoms, and most of that is made up of nucleon mass, which arises as confinement energy in Quantum Chromodynamics~\cite{Kronfeld:2012ym}. Electroweak symmetry breaking is decidedly a minor player.

Only three quark flavors ($u, d, s$) were known when the electroweak theory was formulated. The weak-isospin doublet $\left(\begin{smallmatrix} u \\ d\cos\theta_{\mathrm{C}} + s\sin\theta_\mathrm{C}\end{smallmatrix}\right)_{\mathrm{L}}$, where $\theta_{\mathrm{C}}$ is the Cabibbo angle, captured the known structure of the hadronic charged-current interaction and expressed the universal strength of quark and lepton interactions. Within the Weinberg--Salam framework, however, this single quark doublet gives rise to flavor-changing $s \leftrightarrow d$ neutral-current interactions that are not observed in nature. Glashow, Iliopoulos, and Maiani~\cite{Glashow:1970gm} noted that the unwanted interactions could be cancelled by introducing a second quark doublet, $\left(\begin{smallmatrix} c \\ s\cos\theta_{\mathrm{C}} - d\sin\theta_\mathrm{C}\end{smallmatrix}\right)_{\mathrm{L}}$, involving a new ``charmed'' quark and the orthogonal combination of $d$ and $s$. The absence of flavor-changing neutral currents generalizes to more (complete) quark doublets, and is a striking feature of the experimental data.

To save writing, I have outlined here a theory of a single generation of leptons; the other lepton families are included as simple copies. However, a theory of leptons alone would be inconsistent. In our left-handed world, each doublet of leptons must be accompanied by a color-triplet weak-isospin doublet of quarks, in order that the theory be anomaly free, i.e., that quantum corrections respect the symmetries on which the theory is grounded~\cite{Bouchiat:1972iq}.

Since its invention, the electroweak theory has been supported again and again by new observations, in many cases arising from experiments conceived or reoriented explicitly to test the electroweak theory. 
 I treated this question in some detail in \S3 of ``Unanswered Questions''~\cite{Quigg:2009vq}, to which I refer the reader for specific references. It will suffice here to mention some of the major supporting elements. The first great triumph of the electroweak theory was the discovery of weak neutral currents. This was followed in short order by the discovery of charm (hidden first, then open), which was required in the framework of the electroweak theory, once neutral currents had been observed.  The discovery of the $W$ and $Z$ was the second great triumph of the electroweak theory. Experiment also brought new evidence of richness, including the  the discovery the $\tau$ lepton and evidence for a distinct $\tau$ neutrino and the discovery of the $b$ quark. The top quark completed a third quark generation; the top-quark mass became an essential input to quantum corrections to predictions for precisely measured observatbles. Moreover, finding a third quark generation opened the way to understanding, at least at an operational level, the systematics of \textsf{CP} violation~\cite{Kobayashi:1973fv}. Highly detailed studies at many laboratories confirmed the predictions of the electroweak theory to an extraordinary degree. 
 
 As the electroweak theory emerged as a new law of nature, the question of how the electroweak symmetry was hidden came to center stage. While the default option---the one emphasized in textbooks---was an elementary scalar Higgs boson, electroweak symmetry breaking through some sort of new strong dynamics, or as a message from extra spatial dimensions, or as an emergent phenomenon arising from strong interactions among the weak bosons all received attention from both theory and experiment.

\section{AFTER THE DISCOVERY---OUTLOOK} 
The most succinct summary we can give is that the data from the ATLAS and CMS experiments are developing as if electroweak symmetry is broken spontaneously through the work of elementary scalars, and that the emblem of that mechanism is the standard-model Higgs boson. I refer to Refs.~ \onlinecite{CMSHresults,ATLASHresults,Agashe:2014kda,HiggsPDG14} for details and to Ref. ~\onlinecite{doi:10.1146/annurev-nucl-102313-025603} for perspective.

The bare facts are these: the LHC experiments have found a new unstable particle $H$, with a mass in the neighborhood of $125\gev$. It decays into $\gamma\gamma$, $W^+W^-$, and $Z^0Z^0$ in approximately the proportions expected for a standard-model Higgs boson. The new particle is narrow, for its mass, with the current bounds measured in tens of MeV. The dominant production mechanism has characteristics compatible with gluon fusion through a heavy-quark loop, as foreseen. Topological selections have identified a subsidiary mechanism compatible with vector-boson fusion. Some evidence has been presented for the decays $H \to b\bar{b}$ and $\tau^+\tau^-$. No decays that entail lepton-flavor violation have been observed. The new particle does not have spin 1, studies of decay angular distributions and correlations among decay products strongly favor spin-parity $0^+$ over $0^-$, and while spin 2 has not been excluded in the most general case, that assignment is implausible.

As one measure of the progress the discovery of the Higgs boson represents, let us consider some of the questions I posed before the LHC experiments in Ref.~ \onlinecite{Quigg:2009vq}.
\subsection*{Future Issues (from \textit{``Unanswered Questions \ldots''})}
\begin{description}
\item[1.] What is the agent that hides the electroweak symmetry? Specifically, is there a Higgs boson? Might there be several? \textit{To the best of our knowledge, $H(125)$ displays the characteristics of a standard-model Higgs boson, an elementary scalar. Searches will continue for other particles that may play a role in electroweak symmetry breaking.}

\item[2.]  Is the ``Higgs boson'' elementary or composite? How does the Higgs boson interact with itself? What triggers electroweak symmetry breaking? \textit{We have not yet seen any evidence that $H(125)$ is other than an elementary scalar. Searches for a composite component will continue. The Higgs-boson self-interaction is almost certainly out of the reach of the LHC; it is a very challenging target for future, very-high-energy, accelerators. We don't yet know what triggers electroweak symmetry breaking.}

\item[3.] Does the Higgs boson give mass to fermions, or only to the weak bosons? What sets the masses and mixings of the quarks and leptons? \textit{The experimental evidence suggests that $H(125)$ couples to $t\bar{t}$, $b\bar{b}$, and $\tau^+\tau^-$, so the answer is probably yes. All these are third-generation fermions, so even if the evidence for these couplings becomes increasingly robust, we will want to see evidence that $H$ couples to lighter fermions. The most likely candidate, perhaps in High-Luminosity LHC running, is for the $H\mu\mu$ coupling, which would already show that the third generation is not unique in its relation to $H$. Ultimately, to show that spontaneous symmetry breaking accounts for electron mass, and thus enables compact atoms, we will want to establish the $He\bar{e}$ coupling. That is extraordinarily challenging because of the minute branching fraction.}

%
%
%
%
%

\item[10.] What lessons does electroweak symmetry breaking hold for unified theories of the strong, weak, and electromagnetic interactions? \textit{Establishing that scalar fields drive electroweak symmetry breaking will encourage the already standard practice of using auxiliary scalars to hide the symmetries that underlie unified theories.}
\end{description}

To close, I offer a revised list of questions to build on what our first look at the Higgs boson has taught us. 
\subsection*{Issues Sharpened by the Discovery of \textit{H}(125)}
\begin{enumerate}
\item How closely does $H(125)$ hew to the expectations for a standard-model Higgs boson? Does $H$ have any partners that contribute appreciably to electroweak symmetry breaking?

\item Do the $HZZ$ and $HWW$ couplings indicate that $H(125)$ is solely responsible for electroweak symmetry breaking, or is it only part of the story?


\item Does the Higgs field give mass to fermions beyond the third generation? Does $H(125)$ account quantitatively for the quark and lepton masses? What sets the masses and mixings of the quarks and leptons?

\item What stabilizes the Higgs-boson mass below $1\tev$? 

\item Does the Higgs boson decay to new particles, or \textit{via} new forces?

\item What will be the next symmetry recognized in Nature? Is Nature supersymmetric? Is the electroweak theory part of some larger edifice?

\item Are all the production mechanisms as expected?

\item Is there any role for strong dynamics? Is electroweak symmetry breaking related to gravity through extra spacetime dimensions?

\item What lessons does electroweak symmetry breaking hold for unified theories of the strong, weak, and electromagnetic interactions? 

\item What implications does the value of the $H(125)$ mass have for speculations that go beyond the standard model? \ldots\ for the range of applicability of the electroweak theory?
\end{enumerate}
In the realms of  refined measurements, searches, and theoretical analysis and imagination, great opportunities lie before us!

\subsection*{Summary Points}
\begin{enumerate}
\item The ATLAS and CMS Collaborations, working at CERN's Large Hadron Collider, have discovered a new particle, $H(125\gev)$, that matches the profile of the Higgs boson of the electroweak theory.

\item Observation of decays into the weak bosons, $H \to W^+W^-$ and $H \to Z^0Z^0$, establishes a role for the Higgs boson in hiding the electroweak symmetry and endowing the weak bosons with mass.

\item Evidence for the decays $H \to b\bar{b}$ and $H \to \tau^+\tau^-$, together with characteristics of $H$ production that implicate gluon fusion through a top-quark loop, suggest that the Higgs boson also plays a role in giving mass to the fermions.

\item It will be important to show that $H$ couples to quarks and leptons of the first two generations and to test its role in generating their masses.

\item If the electron mass, in particular, does arise from the vacuum expectation value of the Higgs field, we will have a new understanding of why compact atoms exist, why valence bonding is possible, why liquids and solids can form (cf. \S4.4.2 of Ref. ~\onlinecite{Quigg:2009vq}).

\item The spin-parity of $H$, which is strongly indicated as $0^+$, favors the interpretation as an elementary scalar.

\item Even after its apparent completion by the observation of a light Higgs boson, the electroweak theory raises puzzles. An outstanding question is why the electroweak scale is so much smaller than other plausible physical scales, such as the unification scale and the Planck scale.

\item It is possible that the Higgs boson experiences new forces or decays into hitherto unknown particles.

\end{enumerate}

\section*{DISCLOSURE STATEMENT}
The author is not aware of any affiliations, memberships, funding, or financial holdings that might
be perceived as affecting the objectivity of this review.

\section*{ACKNOWLEDGMENTS}
Fermilab is operated by Fermi Research Alliance, LLC under Contract No.\ DE-AC02-07CH11359 with the United States Department of Energy. I am grateful to Professor Heinrich Meier and the Carl Friedrich von Siemens Foundation for gracious hospitality during the final stages of writing.

I want also to pay my respects to the designers, builders, and operators of the Large Hadron Collider and of the ATLAS and CMS experiments; to the experimental teams for their outstanding achievement; and to the generations of scientists whose work has brought us to the point of savoring the discovery of the Higgs boson.

\appendix
\section{Terms / Definitions\label{sec:terms}}
\begin{description}
\item[Anomaly:] The violation by quantum corrections of a symmetry of the Lagrangian. If anomalies violate gauge symmetry, the theory becomes inconsistent, so the freedom from anomalies becomes a powerful condition on candidate theories.
\item[Charged current:] The weak interaction mediated by the $W^\pm$-boson, first observed in nuclear $\beta$ decay.
\item[Effective field theory:] A description valid over a particular range of energies or distance scales, based on the degrees of freedom most relevant to the phenomena that occur there. Nonlocal interactions mediated by virtual heavy particles are replaced by local interactions that yield the same low-energy limit. The effective theory can only be a valid description
of physics at energies below the masses of the heavy particles, and must be superseded by a more complete (but perhaps still effective theory) on that energy scale.
\item[Flavor-changing neutral current:] A transition that changes quark or lepton flavor, without changing electric charge; strongly inhibited by the GIM mechanism in the standard electroweak theory.
\item[GIM mechanism:] Observation by Glashow, Iliopoulos, and Maiani~\cite{Glashow:1970gm} that flavor-changing neutral-current interactions vanish at tree level, and are strongly inhibited at higher orders, provided that quarks (and leptons) occur in $\mathrm{SU(2)_L}$ doublets. Argued for the necessity of the charm quark.
\item[Goldstone Phenomenon:] The appearance of massless modes whenever a global continuous symmetry of the Lagrangian is broken, in the sense that the vacuum state does not display the full symmetry of the Lagrangian. One massless scalar or pseudoscalar appears for each broken generator of the full symmetry.
\item[Higgs boson:] Elementary scalar particle that is the avatar of electroweak symmetry breaking in the standard electroweak theory, an excitation of the auxiliary scalar fields introduce to contrive a vacuum that does not respect the full \ewgg\ symmetry on which the electroweak theory is built. An unstable particle with mass $125\gev$ that closely fits the profile of the Higgs boson was discovered in experiments at CERN in 2012. 
\item[Lepton:] An elementary (at the present limits of resolution) spin-$\frac{1}{2}$ particle that does not experience the strong interaction. The current roster is composed of three charged particles, $e^-, \mu^-, \tau^-$, and three neutrinos, $\nu_e, \nu_\mu, \nu_\tau$.
\item[Neutral current:] The weak interaction mediated by the $Z^0$-boson, first observed in the reactions $\nu_{\mu} e \to \nu_{\mu}e$ and $\nu_\mu N \to \nu_\mu + \hbox{anything}$.
\item[Quark:] An elementary (at the present limits of resolution) spin-$\frac{1}{2}$ particle that experiences the strong interaction. The current roster is composed of six species: up, down, charm, strange, top, bottom, grouped in three weak-isospin doublets.
\item[Superconductivity:] A phenomenon that occurs in many materials when they are cooled to low temperatures or subjected to high pressure, superconductivity entails zero electrical resistance and the expulsion of magnetic fields.
\end{description}

\section{Abbreviations and Acronyms  \label{sec:abbr}}
\begin{description}
\item[ATLAS:] One of two general-purpose experiments for the Large Hadron Collider, located adjacent to CERN's main campus.  {atlas.ch}.
\item[CERN:] The European Laboratory for Particle Physics straddles the French-Swiss border near Geneva. Its principal research instrument is now the Large Hadron Collider. One of Europe's first common undertakings at its founding in 1954, CERN  now includes twenty-one Member States. {cern.ch}
\item[CMS:] The Compact Muon Solenoid, one of two general-purpose experiments for the Large Hadron Collider. It is located in Cessy, France. {cms.cern.ch}.
\item[LHC:] The Large Hadron Collider at CERN is a two-bore proton synchrotron 27 km.\ in circumference. It is designed to provide proton-proton collisions up to 14-TeV c.m.\ energy and luminosity exceeding $10^{34}\lum$, as well as Pb-Pb and proton-Pb collisions. {lhc.web.cern.ch}.
\end{description}


%
%
\bibliography{HiggsHistX.bib}
\bibliographystyle{ar-style5x.bst}

\end{document}